# Photonic crystal resonator integrated in a microfluidic system

P. S. Nunes,[*] N. A. Mortensen, J. P. Kutter, and K. B. Mogensen

*Department of Micro- and Nanotechnology, Technical University of Denmark, DTU,*

*Building 345 east, 2800 Kgs. Lyngby, Denmark*

[*]*Corresponding author: pedro.nunes@nanotech.dtu.dk*

We report on a novel optofluidic system consisting of a silica-based 1D photonic crystal, integrated planar waveguides and electrically insulated fluidic channels. An array of pillars in a microfluidic channel designed for electrochromatography is used as a resonator for on-column label-free refractive index detection. The resonator was fabricated in a silicon oxynitride platform, to support electroosmotic flow, and operated at $\lambda = 1.55$ μm. Different aqueous solutions of ethanol with refractive indices ranging from n = 1.3330 to 1.3616 were pumped into the column/resonator and the transmission spectra were recorded. Linear shifts of the resonant wavelengths yielded a maximum sensitivity of $\Delta\lambda/\Delta n = 480$ nm/RIU and a minimum difference of $\Delta n = 0.007$ RIU was measured.

Miniaturization of chemical analysis systems has made it possible to work with smaller sample volumes, decrease the analysis time and has introduced portability of analytical devices. To achieve these goals, microfabrication techniques have been used extensively during the last decades in the field of lab-on-a-chip [1]. The detection aspect of lab-on-a-chip systems has, however, received much less attention than the fluidics, since many detection principles are hard to implement and do not scale as favorably, e.g. a reduced interaction length often leads to a compromised sensitivity [2]. Fluorescence detection is therefore often the method of choice for miniaturized systems. However, it typically requires labeling of molecules, thereby adding additional steps to the sample preparation and furthermore altering the chemical properties of the analytes. The change in analyte properties often makes the results difficult to interpret, which is the reason why label-free detection is favored in, e.g., chromatography [3]. Label-free refractive index detection has recently become the working principle of several microfabricated photonic sensors, such as ring resonators [4, 5, 6], Mach-Zehnder interferometers [7], Young interferometers [8] and photonic crystal sensors [9, 10]. Miniaturized separation systems, where label-free detection is preferred, should therefore be able to benefit from these efforts.

In this work, we present a 1D photonic crystal sensor integrated in a microfluidic system. The photonic structure differs in two important aspects from previous work on, e.g., photonic crystal biosensors, which typically consist of an array of holes in a silicon thin film [9]. First of all, using pillars instead of holes makes it possible to pump the liquid in the plane of the device and hence readily integrate the photonic structure within planar microfluidic channels. Secondly, by exploiting dielectric materials for fabrication of the photonic crystal in order to obtain electrically insulated channels, it will be possible for the devices to support electroosmotic flow and thus take advantage of all the know-how developed for electrokinetic separation systems on planar devices during the last decades [11].

The device consists of a 1D array of rectangular pillars etched in a microfluidic channel fabricated in silicon oxynitride (Fig. 1). The pillar array (resonator) has a similar design as used for microfabricated chromatography devices [12, 13].

The fabrication started with an 11 μm thick thermally grown $SiO_2$ used as a bottom cladding layer for the silicon oxynitride single-mode waveguides. A 3.2 μm thick ($\Delta n=0.02$) waveguide layer and a 1.0 μm $SiO_2$ top cladding layer were deposited by plasma enhanced chemical vapor deposition (PECVD) after which they were annealled. A poly-silicon mask was deposited before photolithography and deep reactive ion etching (DRIE) were used to pattern the waveguide layout. A 4.2 μm index-matched borophosphosilicate glass was deposited in the PECVD and later annealed. A second photolithography and etching step was used to pattern the pillars and microfluidic channel. A 200 nm thin layer of amorphous-Si was finally deposited by low pressure chemical vapor deposition (LPCVD). Hermetic sealing of the microfluidic channels was achieved by anodic bonding of a borosilicate glass wafer. The final pattern had a pillar pitch of 5.0 μm, a trench width of 3.0 μm and a depth of 10 μm (Fig. 1). Integrated waveguides for in- and out-coupling of the light were designed with two S-bends in order to assure that only light coupled into the waveguides would be transmitted through the resonator.

The transmission properties of the resonator were tested by pumping fluids with different refractive indices through the microfluidic channel. The measurement setup consisted of a superluminescent diode with a power of 0.2 mW and infrared emission at 1500 – 1600 nm (SLD–76, Superlum, Russia). This broadband light source was connected to an in-fiber polarizer, a polarizer controller, and finally coupled to the chip by the use of single-mode fibers (SMF-28e,

Corning, U.S.). A fiber-coupled spectrometer (Spectro-320, Instrument Systems, Germany) was used for recording of the spectra. The total insertion loss in chips with resonators was 15 dB and 8 dB when coupling light in/out of single mode reference waveguides.

A limitation when using dielectrics as a photonic crystal material is, of course, a lower refractive index contrast and hence a decreased resonance in the structure. The influence of a 200 nm thick a-Si layer on the sidewalls of the resonator was therefore investigated theoretically and experimentally. A thickness of 200 nm has been shown to be thin enough to avoid electrolysis [14], so no bubble formation occurs and electroosmotic flow can be used. A microfluidic channel with nine pillars resembling the fabricated device was simulated using the transfer matrix method [15]. Simulations and experiments were performed with Milli-Q water (n=1.3330) in the microfluidic channel. Introducing a high refractive index material ($n_{Si}$=3.0) in addition to the refractive index of the pillars (n=1.48) leads to an increased finesse [15] and thus to a higher Q-factor, as can be seen from the simulation results (Fig. 2). Another advantage of introducing a-Si on the resonator structure is the occurrence of more resonant peaks allowing to easily tune the resonances to fit to the available light source spectrum. Simulated data yielded a Q-factor of 6 for bare pillars (no a-Si was deposited) and 153 for pillars covered with a-Si. Transmission spectra on microfabricated devices with and without a-Si were measured to verify the simulation results (Fig 2). In the experiments, the Q-factors obtained were 5 and 155 for bare and covered pillars, respectively. These values match the ones expected from the simulations. However, the apparent spectral differences between simulated and experimental results are due to the intensity profile of the broadband light source as well as slight differences in the fabricated pillar dimensions versus the simulation parameters.

Experiments with different ethanol/water solutions (Milli-Q water, n = 1.3330; 10 % (v/v) ethanol in water, n = 1.3395; 25 % (v/v) ethanol in water, n = 1.3507; 50 % (v/v) ethanol in water, n = 1.3616) were used to determine the sensitivity of a resonator with nine pillars (Fig. 1). Aspiration of fluids with a higher refractive index caused the resonance peaks in the spectra to shift to longer wavelengths (Fig. 3a). This shift ($\Delta\lambda$) can - in a small refractive index range - be approximated by a linear fit. The slope from Fig. 3b) shows a sensitivity of 480 nm/RIU for the peak with the highest sensitivity. In the context of the perturbative results presented in [2] this corresponds to a relatively high light-liquid overlap of the order f~40% in comparison to similar devices [4, 9]. The different resonances generally showed sensitivities in the 300-460 nm/RIU range and Q-factors from 70 – 150. So far, the smallest refractive index change measured was $\Delta n$ = 0.007 (S/N=2), for a detection volume of only 0.7 nl, which corresponds to a wavelength shift of 2.15 nm. The high value for the limit of detection is mainly due to the small Q-factor of our resonator (Q=150) [16] and to fluctuations in the signals, since no temperature control of the chips was used during the measurements. The sensitivity was, however, very high, because all light is coupled into the liquid, which results in a much bigger overlap between the optical field and the liquid compared to evanescent based sensors. Simulation results showed that a higher number of pillars would improve the Q-factor of our resonators. In the fabricated devices the non-uniformity of the pillar etching plays a major role and is the reason why no improvement in the Q-factor was observed by increasing the number of pillars. In fact, we believe that another important damping mechanism is the out of plane leakage, since there is no direct light guidance in our resonator [17]. Microring resonators in water [4] can achieve a Q-factor of 1800 and even though the sensitivity is lower, 213 nm/RIU, the limit of detection is better than 0.002 RIU. The sensitivity of our device also compares favorably with what other groups have obtained for planar photonic crystal sensors. Chow *et al*. [9] reported a sensitivity of Δλ/Δn=200 nm/RIU and a detection limit of Δn=0.002 for an ultra-compact 2D photonic crystal sensor, based on an array of circular voids in a silicon thin film. Erickson *et al*. [18] used a single row of holes in a planar photonic crystal configuration that can be used as tunable spectral filter. In this paper, tests performed with several aqueous solutions of $CaCl_2$ showed a sensitivity of Δλ/Δn=100 nm/RIU.

In conclusion, a microfluidic device with integrated refractive index detection based on planar waveguides and a microfabricated pillar array has been presented. The device differs from other chips, because the photonic structure is based on pillars instead of holes to allow liquid transport in the plane of the chip. The device is furthermore hermetically sealed with a glass lid and supports electroosmotic flow, since it is made of glass, thereby extending its range of applications. Future work will focus on increasing the Q-factor by increasing the number of pillars and reducing the channel width through a reduction of the pillar pitch in order to lower the out of plane leakage. Sensitivity, on the other hand, will be improved by increasing the pillars refractive index.

**Acknowledgements**

This work was supported by the Danish Research Council (FTP grant #274-06-0193).

Figures

Fig. 1. Scanning electron microscope image of an on-column 1D photonic crystal refractive index chemical sensor (top view). The resonator pitch was 5.0 μm with a depth of 10 μm. Infrared light was confined in the plane of the chip by 12 μm wide and 3.2 μm thick SiON waveguides.

Fig. 2. Simulated and experimental transmission spectra of a resonator made of nine pillars in a microfluidic channel filled with Milli-Q water with/without a-Si on the pillars.

Fig. 3. (a) Normalized transmission spectra of the 1D photonic crystal microfluidic channel (nine pillars) with four different aqueous solutions of ethanol - 10 %, 25 % and 50 % (v/v) and Milli-Q water. (b) Resonant wavelength plotted as a function of refractive index (n). The error bars refer to fluctuations on the $\lambda_{res}$ after successive injections with different fluids.

Fig. 1.

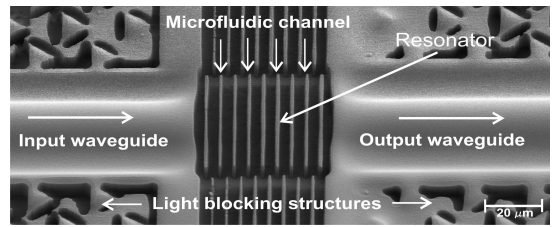

Fig. 2.

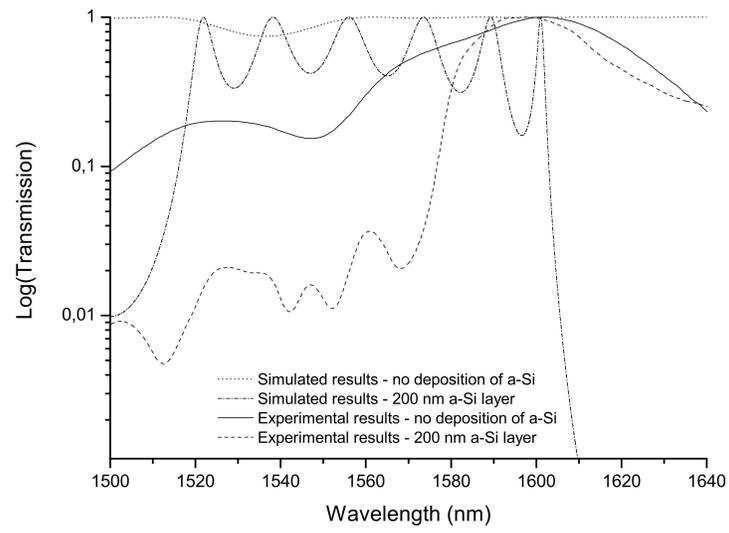

Fig. 3.

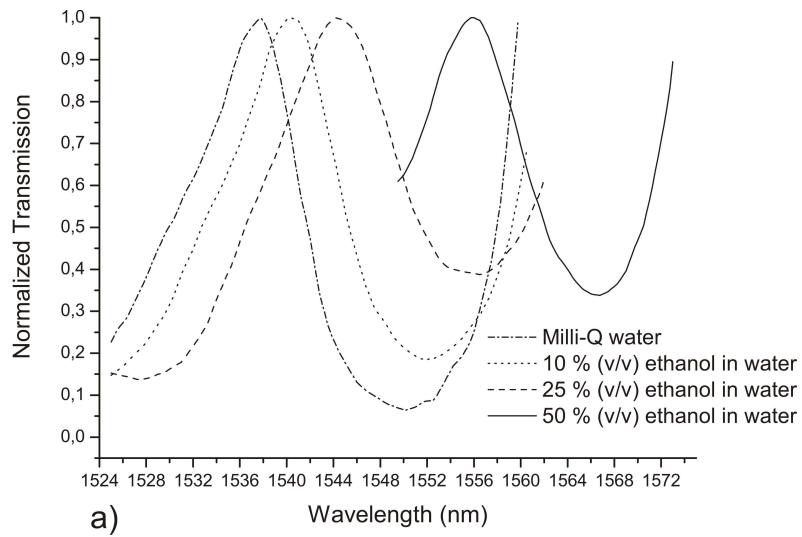

a)

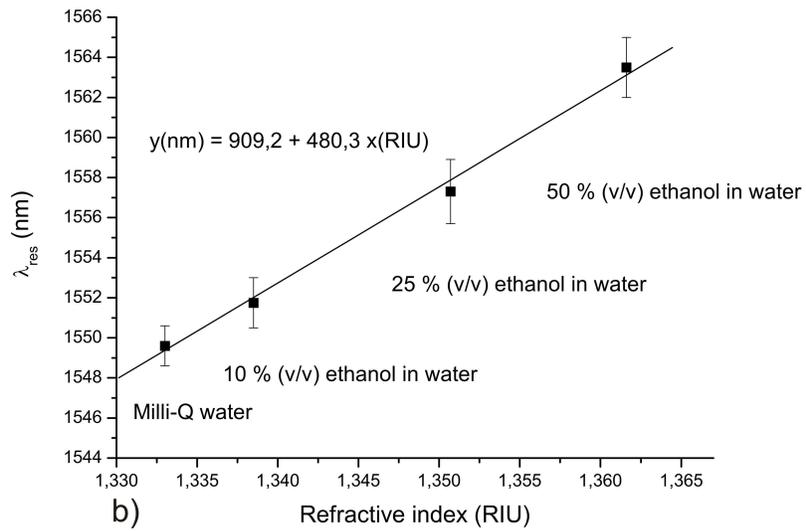

b)